\documentclass[useAMS,usenatbib]{mn2e}
\input psfig.tex
\title[]{The stellar mass fraction and baryon content of galaxy clusters and groups}
\author[S. Andreon]{
S. Andreon,\thanks{stefano.andreon@brera.inaf.it}\\ 
INAF--Osservatorio Astronomico di Brera, Milano, Italy\\
}
\date{Accepted ... Received ...}
\pagerange{\pageref{firstpage}--\pageref{lastpage}}
\pubyear{2009}
\begin{document}
\maketitle

\label{firstpage}

\begin{abstract} 
The analysis of a sample of 52 clusters with precise
and hypothesis-parsimonious measurements of mass, derived from
caustics based on about 208 member velocities per cluster
on average, shows that low mass clusters and groups
are not simple scaled-down versions of their massive cousins
in terms of stellar content: lighter clusters have more
stars per unit cluster mass. The same analysis also
shows that the stellar content of clusters and groups displays an
intrinsic spread at a given cluster mass, i.e. clusters are not similar
each other in the amount of stars they contain, not even at a fixed 
cluster mass.  The stellar mass fraction
depends on halo mass with (logarithmic) slope  $-0.55\pm0.08$ and
with $0.15\pm0.02$ dex of intrinsic scatter at a fixed cluster
mass.  These results are confirmed by adopting
masses derived from velocity dispersion.
The intrinsic scatter at a fixed cluster mass we determine for gas 
mass fractions taken from literature is 
smaller, $0.06\pm0.01$ dex.
The intrinsic scatter in both the stellar 
and gas mass fractions is a distinctive signature that, 
when taken individually, the regions from which 
clusters and groups collected matter, a few tens of Mpc, are yet
not representative, in terms of gas and 
baryon content, of the mean matter content
of the Universe. 
The observed stellar mass fraction values are in marked
disagreement with gasdynamics simulations 
with cooling and
star formation
of clusters and groups. Instead, amplitude and cluster mass dependency
of observed stellar mass fractions are those requested
not to need any AGN feedback to describe gas and stellar mass fractions 
and X-ray scale relations in simple semi-analytic cluster models.
By adding stellar and gas masses and accounting for
the intrinsic variance of both quantities,
we found the the baryon fraction is fairly
constant for clusters and groups with masses between $10^{13.7}$
and  $10^{15.0}$ solar masses and
it is offset from the WMAP-derived value by about 6 sigmas. The 
offset is unlikely to be due to an underestimate of the
stellar mass fraction, and could be related to the possible non
universality of the baryon fraction, pointed out by
our measurements of the intrinsic scatter. 
Our analysis is the first that does not assume
that clusters are identically equal at a given halo
mass and it is also more accurate in many aspects.
The data and code used for
the stochastic computation are distributed with the paper. 
 \end{abstract}


\section{Introduction}

Knowledge of the baryon content of clusters and groups
is a key ingredient in our understanding of the
physics of these objects and in their use as cosmological probes.
In fact, clusters have accreted matter from a 
region of some tens of Mpc, large enough
that their content should be representative of the
mean matter content of the Universe (White et al. 1993).
If this is the case, by measuring the baryon fraction
in clusters, $f_b$, and coupling it with an estimate of $\Omega_b$,
for example from primordial nucleosynthesis arguments or from
CMB anisotropies, gives $\Omega_m=\Omega_b \  f_b$ (e.g. White
et al. 1993; Evrard et al. 1997). Second, the study of how
baryons are distributed in gas and stars and the way this splitting 
depends on halo (cluster or group) mass, should provide clues
to the role played by the various physical mechanisms potentially
active in clusters and groups. 

However, the baryon fraction is far from being fully understood:
WMAP-derived value of the baryon fraction is larger than all values found
in X-ray analysis (i.e. Vikhlinin et al. 2006) even accounting for baryons
in stars (e.g. Gonzalez, Zaritsky, \& Zabludoff 2007),
and gas depletion (e.g. Nagai et al. 2007). X-ray scaling
relations (e.g. halo mass vs Temperature or X-ray luminosity) predicted
on the assumption that the thermal energy of the gas comes solely from the
gravitational collapse are notoriously in disagreement with observed
scalings (e.g. Vikhlinin et al. 2006). Observed and predicted scalings
may be bring in agreement by allowing star formation, and, eventually
a further feedback (e.g. Kravsov et al. 2005, Nagai,
Kravtsov, \& Vikhlinin 2007; Bode et al. 2009; Fabjan et al. 2009). 
In particular, whether a further feedback, i.e. in addition 
to the stellar one, is needed, is largely unknown because
of the uncertainty of the observed stellar mass content of clusters
(e.g. Bode et al. 2009). More generally, recent works on the subject
achieve to reproduce X-ray derived quantities (e.g. baryon fraction or
mass-temperature scaling relations) by basically adding to the cluster model 
a further degree of freedom 
associated with star formation (e.g. Nagai, Kravtsov, \& 
Vikhlinin 2007; Bode et al. 2009; Fabjan et al. 2009), without
adding the corresponding observational constraint, i.e. requiring
that the stellar mass produced in the model fit the data.
We emphasise that gas properties
strongly depend on the amount of stellar mass allowed in the
model (e.g. Nagai \& Kravtsov 2005; Kravtsov et al. 2005; Nagai, 
Kravtsov \& Vikhlinin 2007) and a constraint on the stellar component
has a direct and important consequence on the
gas component of the model.

Several observational determinations of the stellar mass fraction suffer
by important limitations:
published works studied clusters with {\it unmeasured}, or very poorly
measured, masses and {\it unmeasured} reference radii,
while these quantities are requested to be known for the
determination of the stellar mass fraction, as
discussed in later sections.  
It is clear, therefore, that an observational measurement of 
the stellar mass fraction of clusters with known masses and
reference radii is valuable.

The caustic method (Diaferio \& Geller 1997; Diaferio 1999) offers
a robust path to estimating cluster mass and reference radii.
It relies on the identification in projected phase-space (i.e. in
the plane of line-of-sight velocities and projected cluster-centric
radii, $v,R$) of the envelope defining sharp density contrasts
(i.e. caustics) between the cluster and the field region. The
amplitude of such an envelope is a measure of the mass inside $R$.
As opposed to masses derived in other ways 
(e.g. from X-ray, from velocity dispersion,
from the virial theorem, from the Jeans method, etc.) caustic
masses do not require that the cluster is in dynamical
equilibrium (see Rines \& Diaferio 2006 for a discussion). 
There is a good agreement between caustic and lensing masses
for the very few clusters where both measurements are available
(Diaferio, Geller, \& Rines 2005). On larger
cluster samples, caustic masses also shows
a good agreement with virial masses (e.g. Rines \& Diaferio 2006,
Andreon \& Hurn 2010)
and with the extrapolation to larger radii of
dynamical masses derived through a Jean analysis (Biviano \& Girardi 2003). 
Both virial and Jean
masses require, however, 
assume that the cluster is in dynamical equilibrium.

This paper addresses: a) the determination of the stellar
mass fraction in a sample of clusters and groups with well determined
masses and reference radii derived by the caustic method, using,
on average, 208 members per cluster; and b) the determination
of the average baryon content of clusters and groups.

Throughout this paper we assume $\Omega_M=0.3$, $\Omega_\Lambda=0.7$ 
and $H_0=70$ km s$^{-1}$ Mpc$^{-1}$. Magnitudes are quoted in their
native system (quasi-AB for SDSS magnitudes).

\section{Data \& Sample}

The final cluster (halo) sample consist of the 52 clusters 
and groups with accurate caustic
masses (Rines \& Diaferio 2006) fully included in the 
Sloan Digital Sky Survey (hereafter
SDSS) $6^{th}$ data release (Adelman-McCarthy et al., 2008).
Fundamentally, clusters/groups are: a) X-ray flux-selected 
b) with an upper cut at redshift $z=0.1$ (to allow a good caustic
measurement). From the original Rines \& Diaferio (2006) larger 
sample,  we removed (in Andreon \& Hurn 2010)
only clusters at $z<0.03$ to avoid shredding problems (large
galaxies are split in many smaller sources),
two cluster pairs (requiring a deblending algorithm), and
one further cluster, the NGC4325 group, 
because is of very low richness. In the present paper,
one more group, MKW11, has been
removed because the star/galaxy classification of SDSS is
poor at this cluster location, as verified by visual
inspection (see Sec 3.2). 
It turned out also that MKW11 is the halo with smallest
mass in our sample.

We emphasise that only two cluster pairs
have been removed from the original sample because of their
morphology, all the other excluded clusters 
have been removed because they are not fully enclosed
in the sky area observed by SDSS, or have bad SDSS data,
or have suspect masses because
the algorithm used to compute the caustic mass 
converged on a secondary galaxy clump.

The basic data used in our analysis are $g$ and $r$ photometry
from SDSS, down to $r=19$ mag. The latter value is
the value where
the star/galaxy separation becomes uncertain (e.g. Lupton et al. 2002)
and is
well brighter than the SDSS completeness 
limit (e.g. Ivezi{\'c} et al. 2002).  Specifically,
we use petrosian magnitudes for ``total" magnitudes, 
and ``dered" magnitudes  for colours.

\section{Analysis of the individual clusters}

We want to measure
the stellar mass fraction and its dependence on cluster mass.  To
do this, we must a) measure cluster masses and determine reference
radii, b) determine the total luminosity in galaxies, c) estimate
the luminosity of other components (e.g., the brightest cluster
galaxy and intracluster light), and d) convert the stellar
luminosity into stellar mass.
In addition to the above, when the total baryon content is
of interest, we also need the gas mass fraction. 

About point a), we adopted virial masses,
$M200$, and virial radii, $r_{200}$, from 
the caustic analysis of Rines \& Diaferio (2006).
For sake of precision,
$r_{200}$ is the radius within which the enclosed average
mass density is $200$ times the critical density.
Let's consider the remaining points in turn.

\begin{figure}
\centerline{%
\psfig{figure=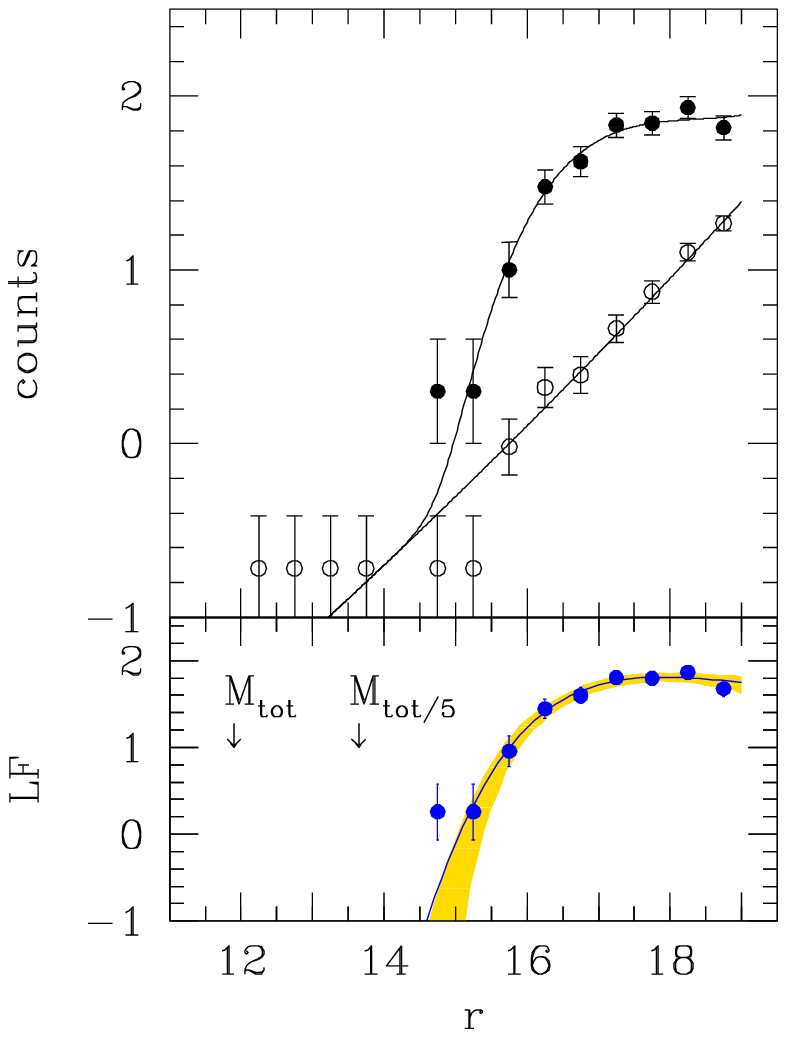,height=6truecm,clip=}%
\psfig{figure=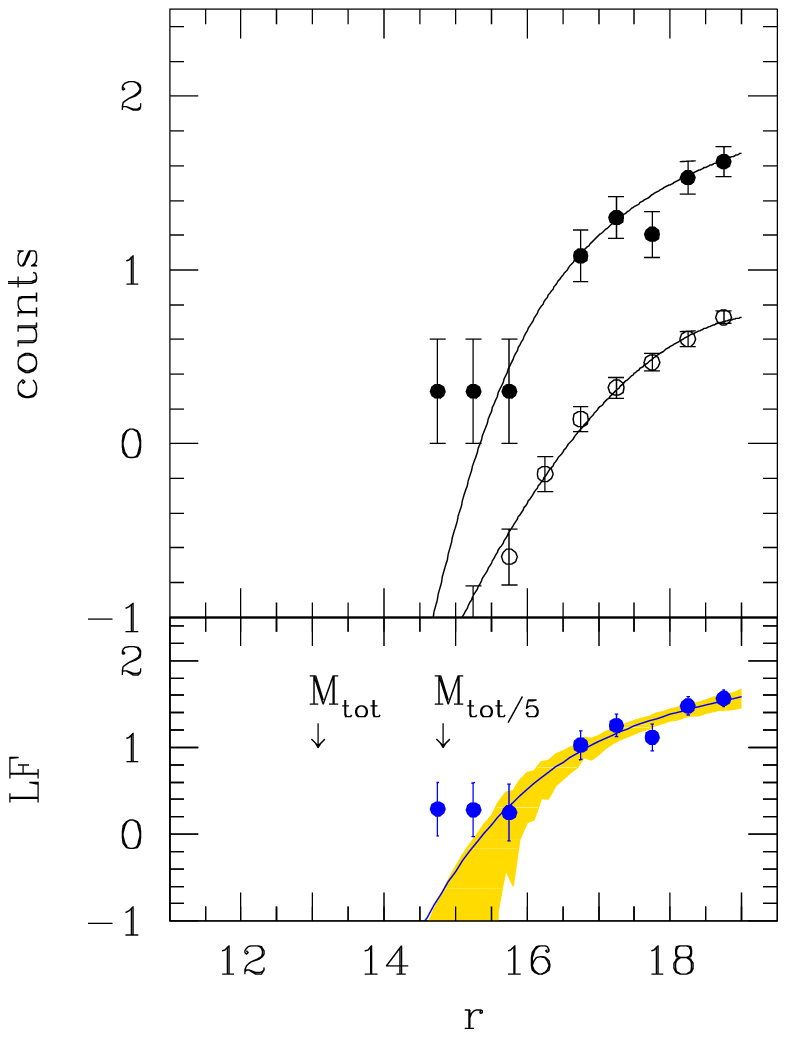,height=6truecm,clip=}%
}
\caption[h]{LF determination. The figure shows galaxy
counts in the cluster direction (upper panel, solid dots),
in a reference line of sight (upper panel, open dots), and
the cluster luminosity function (bottom panel). 
Curves mark the fitted model to un-binned data.
Approximated errors (computed with the usual sum in quadrature)
are marked with bars, precisely computed errors are shaded. The bottom
panel also reports the integral of the luminosity function ($M_{tot}$)
and $1/5^{th}$ of it ($M_{tot/5}$), useful to note the presence of
a bright galaxy or a mis-classified star. The left/right panel refers
to the cluster with the third best/worst stellar mass determination
(Zw1215.1+0400/Abell 954).
}
\end{figure}

\subsection{Luminosity function and its integral}

In order to measure the stellar mass of galaxies, 
we restrict our attention to red galaxies
only: blue galaxies would increase little the stellar mass (e.g. 
Fukugita et al. 1998).  In fact: a) blue galaxies
have lower mass at a given luminosity 
in observations (e.g. Hoekstra et al. 2005) and in
stellar population synthesis models (e.g. Bruzual \& Charlot 2003); and
b) blue galaxies are, on average,
fainter than red galaxies and less abundant in clusters. Therefore
their contribution to the total mass is negligible
(e.g. Fukugita et al. 1998; Girardi et al. 2000) and thus neglected. 
Nevertheless, confirmation of the small role played by blue
galaxies in the total stellar mass budget, perhaps derived
from a better mass tracer such as near--infrared photometry, 
would be valuable especially for less massive clusters where
their contribution is potentially higher in percentage.

In this paper we define red galaxies those 
within 0.1 redward and 0.2 blueward in $g-r$ of the colour--magnitude 
relation, precisely as in Andreon \& Hurn
(2010), and in agreement with works there mentioned.
For the colour center, we took the peak of the colour distribution.
For the slope, we adopted the best fit value derived for the
richest clusters. 
This definition of ``red" is quite simple because for our cluster
sample results hardly
depends on the details of the ``red" definition: the 
determination of the precise location of the colour--magnitude relation 
is irrelevant because the latter is much 
narrower than the adopted 0.3 mag width and because practically all 
galaxies brighter than the adopted luminosity cut are red. 
Colours are corrected for the colour--magnitude 
slope, but the precise slope determination is not
critical given the reduced 
magnitude range explored (less than $\pm 3$ magnitudes)
and the shallow slope of colour--magnitude relation.

The luminosity function
is computed in two different ways: for display purposes only
we bin galaxies in magnitude bins and we account for the background
(galaxies in the cluster line of sight) computing the difference of
counts in the cluster and a reference line of sight
(e.g. Zwicky 1957, Oemler 1974, and many papers since them), the
latter
taken outside the cluster turnaround radius or near to it
for clusters near the SDSS sky boundaries. For display
purpose only, errors are computed with the usual quadrature sum rule.
For our formal analysis, instead, we take a Bayesian approach
as done for other clusters (e.g. Andreon 2006, Andreon et al. 
2006,  2008b, etc):
we use the likelihood given in Andreon, Punzi \& Grado (2005),
which is the extension of the Sandage, Tammann, Yahil (1979)
likelihood to the
case when a background is present. We fit
each cluster, independently on the other ones
and without binning data in magnitude bins. 
We adopt uniform priors for all 
parameters, and we note that any other weak prior would have 
returned a similar result because parameters are well determined
by the data.
For the same reason, a maximum likelihood analysis, 
such as the one advocated in Andreon, Punzi \& Grado (2005), would have
returned identical values for the parameters 
(but with different meanings). However,
the Bayesian approach has a number of advantages, 
amongst which it makes trivial to compute uncertainties
on derived parameters, as the error on the integral of the luminosity
function (that we need in order
to estimate the stellar mass fraction), 
fully accounting for the covariance of all error
terms and with just one
line of code (by typing the about 20 characters in eq. 1 below). 
As usual, all magnitudes
are internally zero-pointed to a number near to the average, 
because this has a number of numerical advantages.

\begin{figure}
\centerline{\psfig{figure=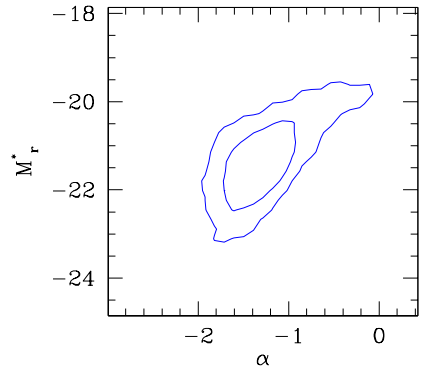,width=6truecm,clip=}}
\caption[h]{Characteristic magnitude $M^*_r$ vs faint
end slope $\alpha$ of Abell 954. There is a clear covariance between
these two Schechter parameters. 68 \% and 95 \% credible
contours are plotted.
}
\end{figure}

\begin{figure}
\centerline{\psfig{figure=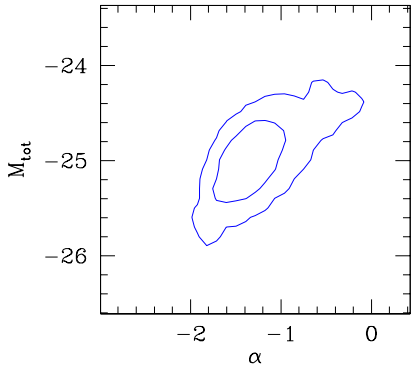,width=6truecm,clip=}}
\caption[h]{Integral of the LF, $M_{tot}$ vs $\alpha$ of Abell 954. 
68 \% and 95 \% credible contours are plotted.}
\end{figure}

Figure 1 exemplifies our analysis for two clusters, chosen as those
having the third best and worst determination of the stellar mass: 
top
panels show galaxy counts in the cluster (solid dots) and
reference (open dots) line of sigh. The cluster contribution is
the excess over the reference line of sight. 
The background is modeled with
a $2^{nd}$ degree power law, the cluster with a Schechter (1976) 
function with the usual parameters $\alpha$ (faint-end slope), $\phi^*$ 
(normalization) and $M^*$ (characteristic magnitude).
The lines show the fitted model on un-binned data. The bottom panel shows the
cluster LF as classically derived (points with the mentioned 
heuristic error bars)
and its Bayesian derivation (mean model with 68 \% confidence
bounds on it, shaded in yellow). 

The analysis is repeated for all 52 (plus one, later discarded) 
studied clusters.

The total luminosity is given by the integral of the luminosity 
function, that, for a Schechter (1976) function is given by: 

\begin{equation}
L_{tot} = \phi^* L^* \Gamma(\alpha+2)
\end{equation}

Figure 2 and 3 show the $M^{\star}_r$ vs $\alpha$    and $M_{tot}$ vs
$\alpha$ of the cluster Abell 954. These figures clarify a number of
technical aspects. First, there is a covariance  between these
quantities. Second, as shown for Abell 954 from the comparison of Figures
2 and 3, the error on $M_{tot}$ is smaller than the quadrature sum of its
parts, and even on $M^*_r$ error alone, owing the covariance between
variables. For our sample of 52 clusters,  the average $M^*_r$ error is
0.45 mag and the average $M_{tot}$ error is 0.21 mag. Third, if,
following almost all previous literature works (e.g. Lin et al. 2003;
Gonzalez et al. 2007; Giodini et al. 2009, etc.), a single value of
$\alpha$ is instead taken (justifying the above by stating that $\alpha$
is  undertermined from the available data), then the derived error would
underestimate the true error on $M_{tot}$ (and even more so the one of
$M^*$). In fact, what literature works measure is the vertical thickness 
at a given $\alpha$, instead than the overall width, obtained  by
projecting the likelihood (posterior) on  on the $y$ axis (i.e.
marginalizing on $\alpha$). 

The luminosity we derived thus far is the one 
emitted from cluster galaxies in a cylinder of radius $r_{200}$.
To compute the stellar mass faction we need instead the luminosity within
a sphere, the latter derived from the luminosity in a cylinder assuming
a Navarro, Frank \& White (1997) distribution with concentration equal to
3. If, instead, the true value of the concentration would be 5, then
we would be under-estimating stellar masses by 0.02 dex, a
negligible quantity 
compared to the final stellar mass uncertainty (0.08 dex,
on average, Sec. 3.4).

\subsection{The bright and faint ends}

The lower panel of Figure 1 is very useful
to detect the presence of galaxies that might give a large contribution
to the total cluster flux, like the
brightest cluster galaxy (BCG, hereafter) 
but also bright galaxies unrelated with the cluster
or mis-classified bright stars. Every galaxy near to, or 
brighter than, $1/5^{th}$ of the
total cluster flux has been carefully checked. Furthermore, 
several fainter
galaxies, down to the magnitude where the (preliminary computed) 
LF predicts less than one galaxy, were also checked.
First of all,
we inspected the SDSS image, and we 
sometime found that the checked galaxy is instead 
a misclassified and saturated star. In such case, the object is
removed from the sample. In such a check, we noted
that the SDSS star/galaxy classification is poor at the
location of the cluster MKW11  (there are many stars
misclassified as galaxies), which has been removed from the sample.
Then, we checked if the candidate BCG is a cluster member or
a foreground galaxy by searching its
redshift in the SDSS and NED archives and comparing it to the
cluster redshift. We either found that the checked galaxy
has a fairly different redshift ($\Delta v > 4000$ km/s)
and, in that case, we removed it from the sample, or very near to it
(less than few hundreds km/s) and we kept it in the sample.
 At this point we have six BCGs much brighter than the LF,
all spectroscopically confirmed as
cluster member.  We now consider the possibility that BGCs are
not drawn from the Schechter (1976) function, at the light
of several literature claims that BCGs may be drawn from a 
different distribution (e.g. Tremaine \& Richstone 1977). 
In order to 
caution us against the risk of missing this source of 
stellar mass,  
we (temporary) remove the object from the sample, 
or better, we remove a magnitude range largely including the
BCG, and
we re-compute the LF rigorously accounting for missing luminosity range
(censored and truncated observations, e.g. in Andreon, Punzi \& Grado 2005). 
We re-integrate the model LF over the full luminosity range, and
we add back the temporary removed BCG flux. We find that
the median flux of the six BCGs is 16 per cent of the 
cluster flux.

It is well known that shallow photometric data miss
the flux coming from the galaxy outer regions (e.g. Andreon 2002),
or, equivalently, that Petrosian magnitudes listed in the SDSS
catalog
underestimate the total galaxy flux  (e.g. Blanton et al. 2001). For
de Vaucouleurs (1948) profiles, typical of red galaxies of interest
here, 
Petrosian magnitudes underestimate the total flux by
about 15 \% (Blanton et al. 2001) for galaxies with the size of
those studied in this paper. 
Our total flux is corrected for this missed flux. 

There is one more component to be accounted for, the intracluster light.
Of course, it should be counted only once in our measurement of
the total flux. Therefore, its value should not include the light coming
from the galaxy outer halos, from the BCG, and 
from faint galaxies (e.g. too faint to be individually detected) because we
already accounted for these three terms. Zibetti et al. (2005)
measure it by accounting for the three mentioned terms 
on a stack of clusters and found a small (10 \% within
500 kpc, about $r_{200}/2$ for their clusters) and decreasing fraction with 
clustercentric radii. At 
the radius of interest, $r_{200}$, it is a minor term, and therefore 
it is neglected. The small spatial extent of the
ICL is confirmed by the Gonzalez et al. (2007)
analysis: 80 \% of the BCG+ICL total light is contained in the
inner 300 kpc.

We can independently confirm the smallness of the ICL luminosity
using measurements from Gonzalez et al. (2007), after accounting
for different definitions of ICL among works. These
authors measured the intracluster+BCG light and found
that 30 \% of the total light is in the intracluster+BCG 
light at $r_{200}$. These authors studied clusters that
contain a
dominant BCG. We estimate
the contribution of the BCG light in Gonzalez et al. (2007) sample
as similar to the one in clusters dominated by
a BCG in our own sample, about 16 \%. Gonzalez et al. (2007)
quote that a few more per cent of the faint galaxy flux,
we counted with LF, ends up in the their BCG+ICL measurement, and
we comment that some few more per cent of the flux from the 
galaxy halo also likely ends up in the their BCG+ICL measurement.
In summary, the ICL, defined as in our own paper, measured
by Gonzalez et al. (2007) is 10 \% with large errors, 
because of the indirect path used to infer it.
This estimate confirms the measurement performed
by Zibetti et al. (2005): the ICL (as defined in our own and
Zibetti et al. 2005 paper) contribution is negligible at $r_{200}$.
 We emphasize that some other papers uses the term ``cD halo"
to indicate the flux that is counted with/as
``intracluster light".

\begin{table}
\caption{Stellar masses and errors}
{
\scriptsize
\begin{tabular}{l l l l}
\hline
id & $\log M_{\star}$ & id & $\log M_{\star}$ \\
  & [$M_{\odot}$] &   & [$M_{\odot}$]  \\
\hline
A0160          & $12.41\pm 0.10$ & A1728	  & $12.52\pm 0.09$\\
A0602          & $12.49\pm 0.09$ & RXJ1326.2+0013 & $12.27\pm 0.13$\\
A0671          & $12.74\pm 0.08$ & A1750	  & $12.84\pm 0.07$\\
A0779          & $12.32\pm 0.09$ & A1767	  & $12.87\pm 0.06$\\
A0957          & $12.46\pm 0.09$ & A1773	  & $12.78\pm 0.07$\\
A0954          & $12.56\pm 0.13$ & RXCJ1351.7+4622& $12.32\pm 0.14$\\
A0971          & $12.75\pm 0.07$ & A1809	  & $12.80\pm 0.08$\\
RXCJ1022.0+3830& $12.46\pm 0.10$ & A1885	  & $12.47\pm 0.12$\\
A1066          & $12.82\pm 0.06$ & MKW8 	  & $12.23\pm 0.12$\\
RXJ1053.7+5450 & $12.64\pm 0.07$ & A2064	  & $12.34\pm 0.08$\\
A1142          & $12.48\pm 0.10$ & A2061	  & $12.96\pm 0.06$\\
A1173          & $12.34\pm 0.11$ & A2067	  & $12.44\pm 0.09$\\
A1190          & $12.70\pm 0.06$ & A2110	  & $12.53\pm 0.09$\\
A1205          & $12.64\pm 0.09$ & A2124	  & $12.80\pm 0.06$\\
RXCJ1115.5+5426& $12.60\pm 0.07$ & A2142	  & $13.13\pm 0.04$\\
SHK352         & $12.58\pm 0.11$ & NGC6107	  & $12.48\pm 0.08$\\
A1314          & $12.53\pm 0.16$ & A2175	  & $12.78\pm 0.07$\\
A1377          & $12.66\pm 0.08$ & A2197	  & $12.53\pm 0.08$\\
A1424          & $12.74\pm 0.07$ & A2199	  & $12.84\pm 0.05$\\
A1436          & $12.71\pm 0.07$ & A2245	  & $12.96\pm 0.06$\\
MKW4           & $12.47\pm 0.14$ & A2244	  & $13.03\pm 0.07$\\
RXCJ1210.3+0523& $12.45\pm 0.09$ & A2255	  & $13.22\pm 0.04$\\
Zw1215.1+0400  & $12.85\pm 0.05$ & NGC6338	  & $12.39\pm 0.11$\\
A1552          & $12.90\pm 0.07$ & A2399	  & $12.69\pm 0.07$\\
A1663          & $12.77\pm 0.07$ & A2428	  & $12.43\pm 0.11$\\
MS1306         & $12.28\pm 0.10$ & A2670	  & $12.96\pm 0.05$\\
 \hline 	
\end{tabular}
}  
\end{table}    

\subsection{The luminosity to stellar mass conversion}

For the luminosity to stellar mass conversion we
adopt the $M/L$ value derived by
Cappellari et al. (2006). As in
previous works, we assume that in the galaxy regions
studied by Cappellari et al. (2006) the contribution of
non-stellar matter is negligible.

The data of Cappellari et al. (2006) are also consistent
with up to 30 \% non-stellar matter. If it is 15 \% on average,
then stellar massess would be 0.07 dex lower. If instead
stellar mass were derived assuming an old single stellar population
of solar metallicity and Chabrier initial mass function, we would
find only 0.1 dex lower values.

\begin{figure}
\psfig{figure=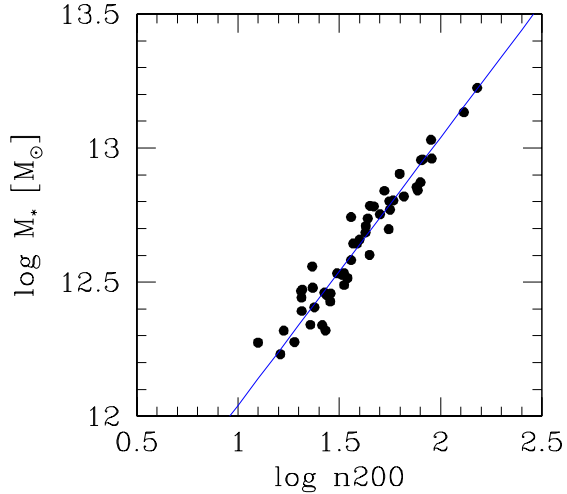,width=8truecm,clip=}
\caption[h]{Richness vs stellar mass. There is tight relation
between the number of bright red galaxies, $n200$, and the total
stellar mass, $M_{\star}$, in clusters. The line marks a relation
with slope one.
}
\end{figure}

\subsection{Results}

Table 1 gives the found stellar masses within $r_{200}$ and their
errors.  Stellar mass errors are small in absolute terms,
0.08 dex on average, and also smaller than
halo mass errors (i.e. errors from caustics, 0.14 dex on average).

Figure 4 plots the stellar mass, derived in the present paper,
against the cluster richness $n_{200}$, i.e.
the number of red galaxies brighter than $M_V=-20.0$ mag, derived
for the very same sample by Andreon \& Hurn (2010). There is
a good agreement between the two quantities, which are
basically two ways to summarise the 
luminosity content of clusters (by counting galaxies or
photons): 

$ lgM_{\star} = (\log n200 -2)  + 13.04 \pm 0.03 $

The slope of the regression is fixed to 1.0 (it is not
a fit to the data) and the quoted error is just the formal
error. For our sample, the two quantities, $lgM_{\star}$ and 
$\log n200$ are
known with the same amount of precision (0.08 dex).

The good agreement between the two summaries of the cluster luminosity
content is a confirmation of the correctness
of the two derivations. We note, in fact, that the present
derivation uses a more constrained model (a shape for the luminosity 
function and for background galaxy counts), and
more (i.e. also fainter) data than the 
derivation of $n_{200}$ in Andreon \& Hurn (2010). The
proportionality of stellar mass and richness is in agreement with
the very small, if any, dependency of the 
faint end slope of the luminosity function 
with richness (e.g. Garilli, Maccagni \& Andreon 1999; Paolillo
et al. 2001; Andreon 2004), 
and with the direct
determination of Rines et al. (2004) from a small sample
of 9 clusters. These authors consider, however, $lgM_{\star}$ and
$\log n200$ integrated down the same limiting magnitude,
differently from our choice.

\subsection{Comparison with literature}

Our stellar mass determination fundamentally differs from previous 
published works from three major
points of view: a) in the way we choose the clusters to be studied;
b) in the adopted reference radius;  
and c) in the performed analysis.

a) First and foremost, if a precise stellar mass within a reference radius of
clusters have to be derived, it is strongly preferable 
(not to say essential) that:
 
i) the studied clusters are truly existing objects. 
Our clusters
and groups are truly existing objects, with an extended X-ray emission
and, on average, 208 spectroscopically confirmed members. Most
of Giodini
et al. (2009) systems are noisy X-ray detections which are ambiguous both
in terms of detection and extent, ``cleaned" by asking a
spatial matching with an overdensity of galaxies to decrease contamination
by point sources and blends of point sources misclassified as extended 
X-ray sources (Finoguenov et al.
2008). Only half of the 'surviving' detections have three or more
concordant redshifts (Giodini et al. 2009), and only a minority
is currently spectroscopic confirmed, according
to Gal et al. (2008), that show that three concordant redshifts
occur frequently by chance in real redshift surveys (a further
example is given in Sec 3.6 of Andreon et al. 2009 using the VVDS
survey).

ii) the reference radius in which stellar masses have to be
measured is known and individually measured.
All our clusters have individually measured radii (by Rines
and Diaferio 2006). Clusters in
Gonzalez et al. (2007) and Giodini et al. (2009) 
have radii inferred from X-ray scaling relations assuming that
these are scatter-free, contrary to observations
(e.g. e.g. Stanek et al. 2006; Vikhlinin et al. 2009;
Andreon \& Hurn 2010). In other terms, these works assume that
the radius appropriate for the studied clusters is the one of an average
cluster having the same observable (e.g. X-ray flux), ignoring
that clusters have a variety
of radius values, even restricting the attention to those 
with a given value of the observable (say X-ray luminosity, it is just
enough to think to cool-core and not cool-core clusters). 
In passing, the scaling adopted by 
Gonzalez et al. (2007) is said in agreement with Hansen et al. (2005),
which is now known to return radii wrong by a factor two (Sheldon et al.
2007, Rykoff et al. 2008).

iii) the studied clusters are located in a narrow range of redshift, in
order not to be obliged to assume an evolution on cluster scaling relation (e.g. 
how the $L_X$-mass scaling evolves) or on M/L.
All our clusters are in the local universe ($z<0.1$), saving
us from making an hypothesis on how to relate 
parameters (masses, virial radii,
etc.) measured at widely different different redshifts. Other  
works (e.g. Giodini et al. 2009) consider
objects in a large redshift range (e.g. $0.1<z<1$) and neglect
the uncertainty on evolution.

b) We believe our adopted radius, $r_{200}$, a
better choice for the determination of stellar masses
than the radius adopted by other authors, $r_{500}$. 
First, $r_{500}$ is small
enough that the stellar mass within $r_{500}$ depends on the precise
definition of ``center" (barycenter, BCG location, X-ray peak, etc).
If the center is measured with a finite degree of accuracy
(which is often the case), a small radius leads to a systematic
underestimate of the stellar mass because of centroiding errors. If instead
one pretends that the cluster center is coincident with the BCG 
location, then a systematic overestimate is introduced,
because the observationally derived value will be boosted
by the presence of the BCG. Systematics
are strongly reduced if $r_{200}$ is used, as it encloses most of  the
cluster. The $r_{500}$ radius also makes an under-optimal use of the
optical data, stellar masses have smaller
error if  $r_{200}$ is used in place of $r_{500}$, as
it is fairly obvious ($r_{500}$ is only a few times the BCG size),
and as we verified for our sample. Furthermore, the choice of a larger 
radius makes
the BCG and intracluster light contribution  
small in percentage and thus their precise contributions largely irrelevant
for the aim of determining the total amount of mass in stars (and in
baryons).
A larger than $r_{500}$ radius is also what is needed to compare
with theory, because the latter has big difficulties in predicting the stellar
mass fraction on such small scale. On the other end, the use of a radius
larger than $r_{500}$ makes more complicate to compute the total baryon
mass fraction, being the gas mass fraction usually measured at smaller 
radii.

c) For what concerns the analysis,  the way we derived the integral
of the luminosity function is rigorous: 
the use of the likelihood function for unbinned data
is a significant improvement (Cash 1976; Sandage, Tammann \& Yahil
1979; Andreon, Punzi \& Grado  2005; Humphrey, Liu, \& Buote 2009) 
above previous approaches
that bin data, and even more above those that
use simplified schemes, as quadrature sums, to
combine errors (see Andreon, Punzi \& Grado  2005 for details).   
Furthermore, differently from previous works, we do not identify
the luminosity function, which is a positively defined quantity,
with the difference of two galaxy counts, that may be negative. Our
choice of marginalizing over the LF parameters is in agreement
with axioms of probability and logic. Keeping some of them,
e.g. $\alpha$ fixed, as in most literature papers,
contradicts them. Finally, and differently from all 
other works, we allow each cluster to have
its own faint end slope and characteristic magnitude, given
that the luminosity function differs from cluster to cluster 
(e.g. Virgo and Coma clusters: Bingelli, Sandage, 
\& Tammann 1988).

\begin{figure*}
\centerline{%
\psfig{figure=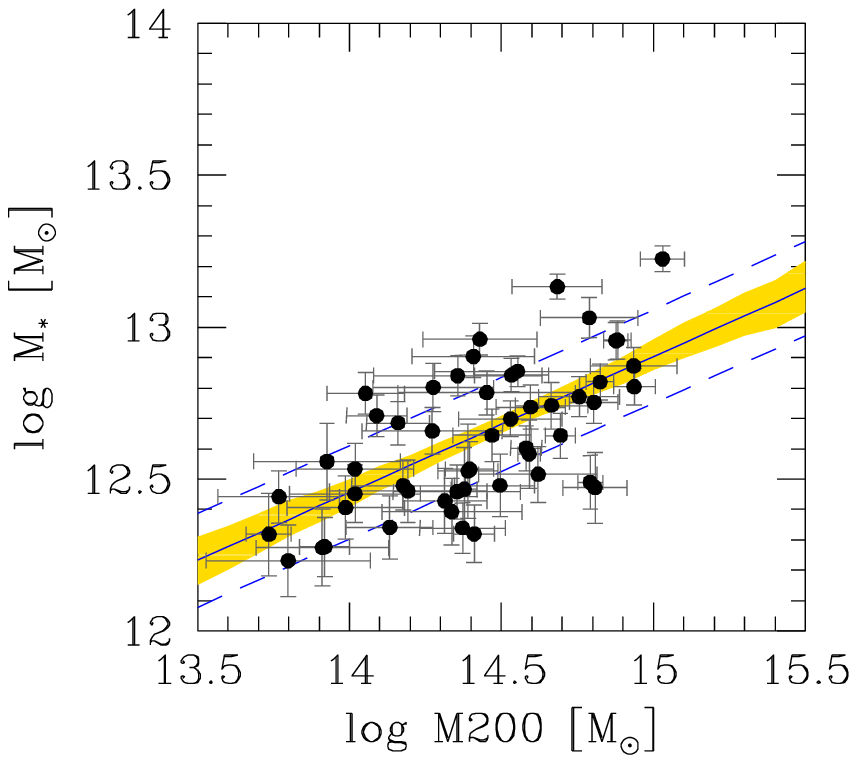,width=8truecm,clip=}
\psfig{figure=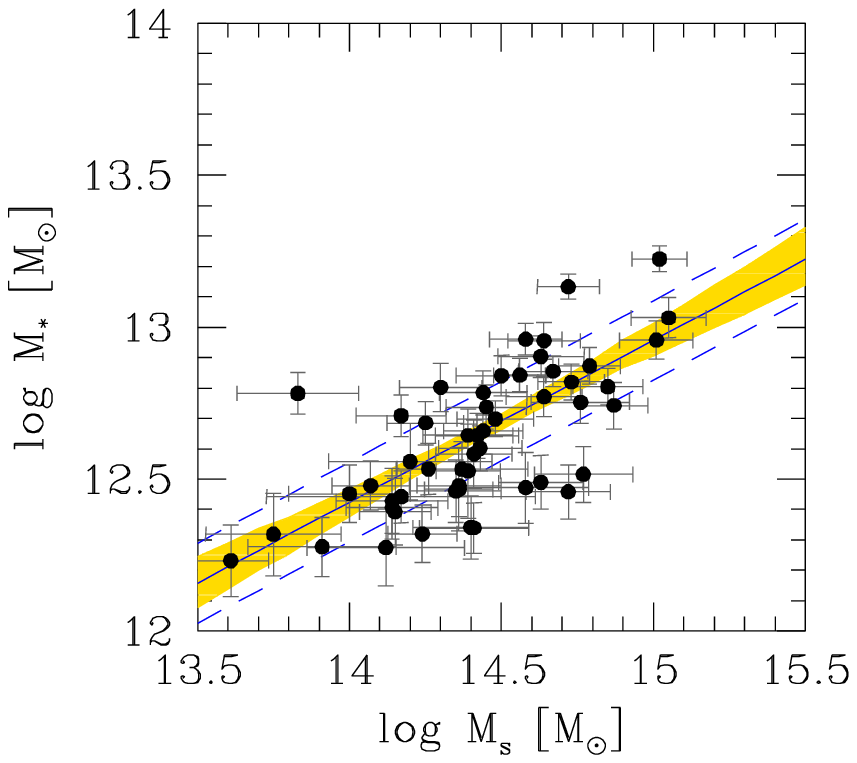,width=8truecm,clip=}}
\caption[h]{Stellar mass vs cluster mass scaling. 
The solid line marks 
The mean relation between stellar mass and halo mass is marked with
a solid line, its 68 \% uncertainty is shaded (in yellow). 
The dashed lines
show the mean relation plus or minus the intrinsic scatter $\sigma_{scat}$. 
Error bars on the data points represent observed
errors for both variables. The distances between the data and the regression line is due in part to the
observational error on the data and in part to the intrinsic scatter.
Left panel uses caustic masses, right panel uses velocity-dispersion based
masses fixed by numerical simulations.}
\label{fig:fig1}
\end{figure*}

\begin{figure*}
\psfig{figure=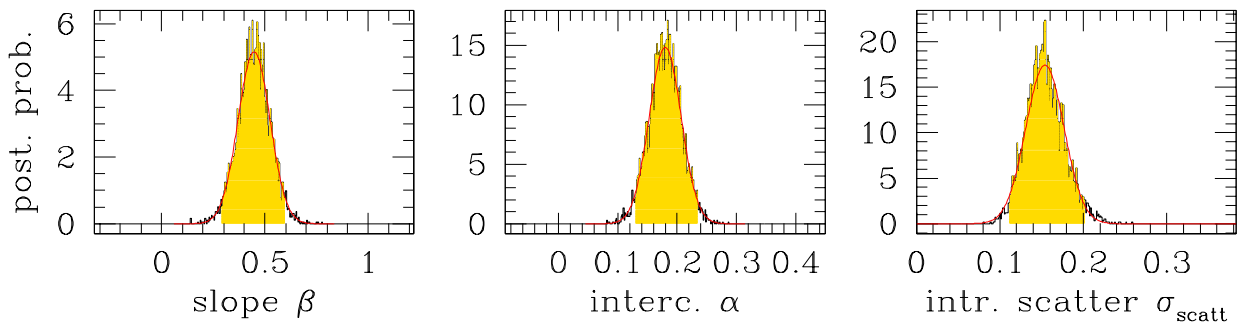,width=17truecm,clip=}
\caption[h]{Posterior probability distribution for the
parameters of the halo mass-stellar mass scaling.
The black jagged histogram shows the posterior as computed
by MCMC, marginalised over the other parameters. The red curve
is a Gauss
approximation of it.  The shaded (yellow) range shows
the 95 \% highest posterior credible interval. 
}
\end{figure*}

\section{Collective analysis of the whole sample}

\subsection{Stellar mass vs halo mass}

In order to fit the trend between stellar and total mass
we use the statistical model (fit) detailed in the Appendix A.
Essentially, our model assumes that the true stellar mass 
and true halo mass are linearly related with some intrinsic scatter 
but rather than having these true 
values we have noisy measurements of both stellar mass and halo mass,
with noise amplitude different from point to point.
In the statistics literature, such a model is know as an 
``errors-in-variables regression'' (Dellaportas \& Stephens, 1995)
and has been widely used before, including more
complex situations (e.g. Andreon 2006, 2008; 
Andreon et al. 2006, 2008a, 2008b, Andreon \& Hurn 
2010, Kelly 2007,  etc.). 
The model is fully specified, and the code listed, in Appendix A.

Using the (fitting) model above, we found, for our sample of 52 clusters:

\begin{equation}
lgM_{\star} =  (0.45\pm0.08)\  (\log M200-14.5) +12.68\pm0.03
\end{equation}
(Unless otherwise stated, results of the statistical computations 
are quoted in the form $x\pm y$ where $x$
is the posterior mean and $y$ is the posterior standard deviation.)

The left panel of Figure 5 shows the relation between stellar mass and halo mass, observed
data, the mean scaling (solid line) and its 68 \% uncertainty (shaded yellow
region) and the mean intrinsic scatter (dashed lines) around
the mean relation. The $\pm 1$ intrinsic scatter band is 
not expected to contain 68 \% of the data points, because of 
the presence of measurement errors.

Figure 6 shows the posterior marginals for
the parameters: slope, intercept, and intrinsic
scatter $\sigma_{scat}$. These marginals
are well approximated by Gaussians.

The slope is very different from one, i.e. low mass clusters
are not simple scaled-down version of high mass clusters: they
have more stars per unit halo mass than their more massive
cousins,  in agreement with Girardi et al. (2000), and
other works. Equivalently, the stellar mass fraction 
decreases with increasing stellar mass,
as better shown in sec 4.3.

The intrinsic stellar mass scatter at a given halo mass, 
$\sigma_{scat}=\sigma_{lgM_{\star}|lgM200}$, is $0.15\pm0.02$ dex.
This is the intrinsic scatter, i.e. the term left after
accounting for measurement errors. It is clearly non-zero (see right panel
of Figure 6).
This is a sort of ``cosmic variance": at a given halo mass, 
clusters are not all equal in terms of the amount of stars they have,
but show a spread of stellar masses.  Alternatively,
the intrinsic scatter is a manifestation of an
underestimate of the errors. This is unlikely to be the case, 
because an intrinsic scatter is seen also
in gas masses and in
numerical simulations of star and gas masses (later discussed),
and therefore
we need that errors on four different observables (observed 
gas and stellar masses, gas and stellar masses predicted in
numerical simulations)
are underestimated, which is unlikely. Next section addresses in
detail a further hypothetical possibility, whether the intrinsic scatter of 
stellar masses is due to underestimated errors on caustic masses
(underestimate that, even if present, does not explain anyway why a scatter 
is also seen in numerical simulations).
No matter which is the source of the intrinsic scatter, its
presence has a few consequences: first, 
larger cluster samples
are requested to measure the average
stellar mass fraction with a given precision. Second, and conversely, 
the existence of a non-zero intrinsic scatter 
is a technical complication to be accounted for
in the determination of the trend between stellar mass
(or stellar mass fraction) vs cluster mass. The intrinsic scatter is,
as mentioned, rigorously accounted for in our fitting model.

\begin{figure}
\psfig{figure=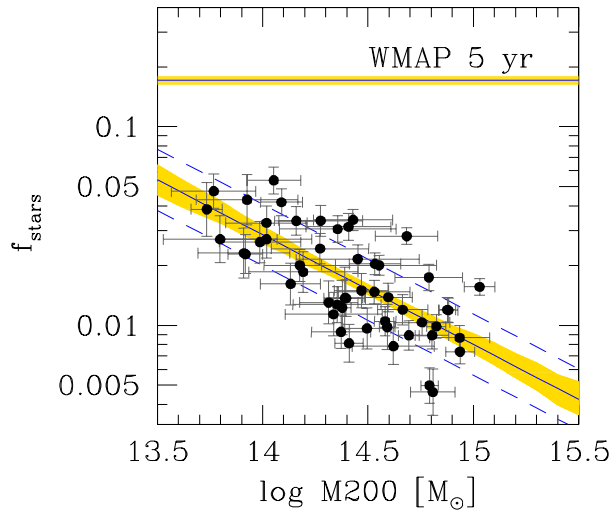,width=8truecm,clip=}
\caption[h]{Stellar mass fraction.
Symbols as in Figure 5. The fit is performed in the stellar vs total mass plane.
The distances between the data and the regression line is due in part to the
observational error on the data and in part to the intrinsic scatter. The WMAP 5 yr value and errors (Dunkley et al. 2009) are marked.}
\end{figure}

\subsection{Checking caustic masses}

Because of the relative novelty of caustic masses, and the
hypothetical possibility that the intrinsic scatter on stellar
masses might be due to an underestimate of caustic mass errors, 
we now
replace caustic mass by a mass, $M_s$, derived from velocity
dispersion using a relation calibrated with numerical simulations
in Biviano et al. (2006). As shown in Andreon \& Hurn (2010),
the mass derived using the calibration in Evrard et al. (2008)
gives almost indistinguishable numbers, because the
two calibrations are almost identical for our clusters. Velocity
dispersions are taken from Rines \& Diaferio (2006).

To use the masses $M_s$ in place of the caustic ones, 
we need only write their values (and their errors) in the data file and 
run our fitting model, listed in Appendix. Mass errors are derived
by combining in quadrature velocity dispersion errors (converted
in mass) and 
the intrinsic noisiness of $M_s$ (12 \%, from Biviano et al. 2006).
We found:

\begin{equation}
lgM_{\star} =  (0.53\pm0.08)\  (\log M_s-14.5) +12.69\pm0.03
\end{equation}

with an intrinsic scatter of $0.13\pm0.03$. By changing the 
source of halo masses, regression parameters 
(slope, intercept and intrinsic
scatter) do not change.
Therefore, the observed intrinsic scatter cannot be due
to (unknown) systematics of caustic masses.
Data and fit for velocity dispersion derived masses
are shown in the right panel of Fig 5.

The insensitivity of our results to which mass is used 
is expected, because 
in Andreon \& Hurn (2010) we show the absence of a gross offset
or tilt between caustic and velocity-dispersion based masses, and
that the error quoted for the caustic mass is as precise as (or as 
wrong as) the error quoted for the velocity dispersion-based mass. 

To summarize, the scatter on the amount of stars that clusters contain
at a given halo mass is not due to an un-accounted systematic 
of the halo mass.

\subsection{Stellar mass fraction}
 
Figure 7 plots the stellar mass fraction vs the cluster mass. The fraction
is computed following its definition: it is the logarithmic
difference of the stellar mass, $M_{\star}$, and the halo mass,
$M200$. The fit
has been performed in the stellar mass
vs halo mass plane, and so derived errors are
shaded in Figure 7. In addition to the exact
analysis (marked with lines and shadings), we also report
approximated errors, marked as error bars, based on just 
the error on stellar
mass, for simplicity. Our fit to the data
accounts
for the intrinsic scatter, and also simply solves a further 
problem that affects previous analysis:
the fit in the fraction vs halo mass plane performed by 
other authors (e.g. Gonzalez et al. 2007; Giodini et al. 2009) has the
halo mass both on the abscissa and in the ordinate (it is at the 
denominator of the fraction). As a consequence,
the values fitted by other works have correlated errors. 
However, past works used 
fitting methods that assume errors to be uncorrelated. 
Our solution, fitting in
the stellar mass vs halo mass, where measurement
errors are uncorrelated, 
solves this problem too.

The decrease
in the stellar mass fraction is stunning, with a
slope equals to $-1+0.45\pm0.08=-0.55\pm0.08$. The quality
of this result can be
better appreciated after noting that the trend above was
considered not constrained by data in recent papers: Allen et al. 
(2008) assume a stellar mass fraction
proportional to the gas fraction (when instead
the two fractions have opposite dependencies with halo mass, as 
shown in Figure 8 and discussed in Sec 4.4), whereas Ettori et al. (2009) 
considered a number of recipes, because of lack of conclusive
data. Bode et al. (2009)
consider the slope as a free (i.e. not constrained by any
stellar fraction observation) parameter.

The
value of the slope is robust to systematic errors
affecting the conversion factor from luminosity
to stellar mass. In fact, to bias the found slope,
we need that the $M_{\star}/L$ value of the galaxies
depends on the mass of the cluster. It is difficult
to imagine why the initial stellar mass function 
(which largely regulate $M_{\star}/L$ value) should be
different in galaxies that are 
in clusters of different masses. Furthermore, the $M/L$
values in Capellari et al. (2006) comes from galaxies lying in
halos of different masses. Finally, the fundamental plane
shown no halo-mass or environmental dependency (Pahre et al.
1998).

\begin{figure*}
\centerline{%
\psfig{figure=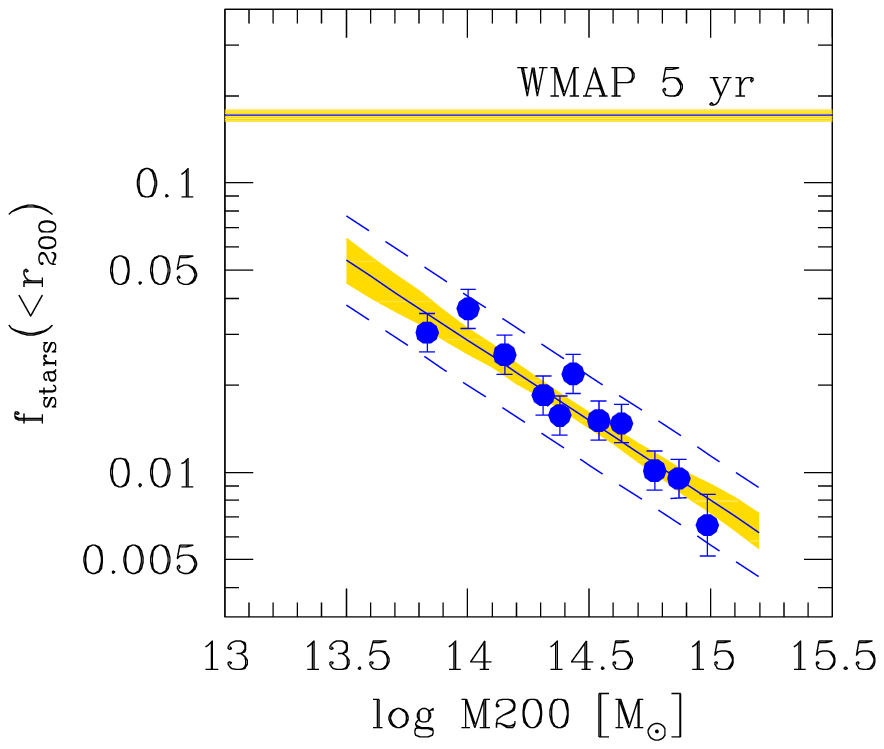,width=8truecm,clip=}
\psfig{figure=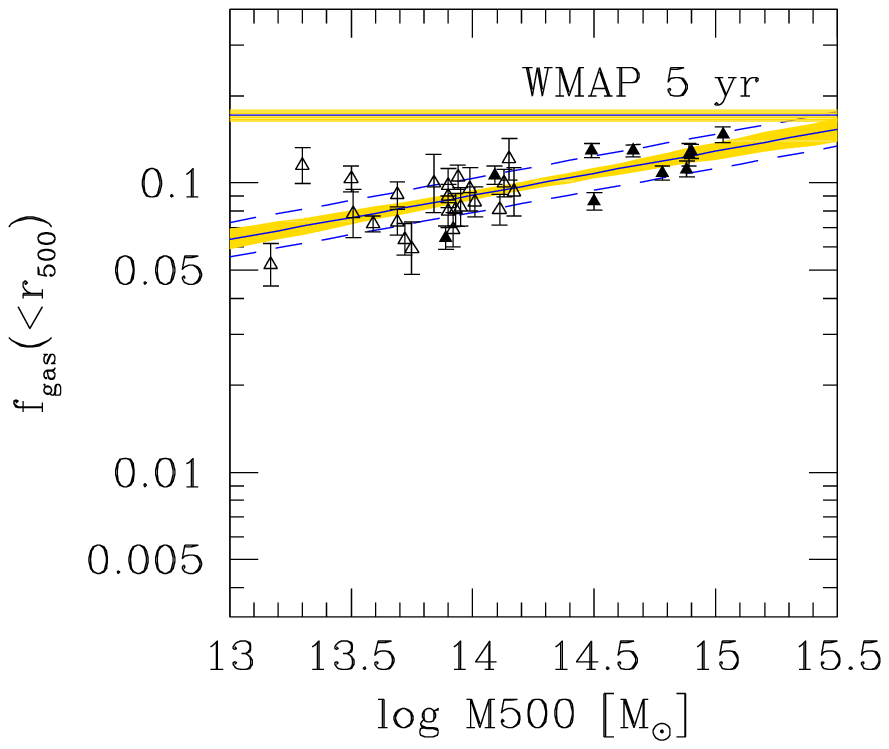,width=8truecm,clip=}}
\caption[h]{Stellar and gas mass fraction.
{\it Left panel:}
Closed (large, blue) circles mark the stellar mass fraction in stacks of 5
clusters each, with the exception of the most massive point, 
composed by just two clusters. Error bars mark approximate errors.
The solid line and shaded region marks the mean model and its (rigorous) 68 \% confidence
error, fitted on individual data points in the  stellar vs total mass plane.
{\it Right panel:}
Open/solid triangles
mark the gas mass fraction from Sun et al. (2009) and Vikhlinin et al.
(2006), respectively. The solid line and shaded region mark the mean model and 
its (rigorous) 68 \% confidence error, derived by us using their
data.
The WMAP 5 yr value and errors (Dunkley et al. 2009) are marked in both
panel.}
\end{figure*}

The left panel of Figure 8 shows the fraction of mass in 
stars, but after stacking clusters
in bin of five clusters each,
with the exception of the highest mass bin, composed
by two clusters only. The figure reports both the 
rigorous computation of errors (shading),
computed in the stellar vs total mass plane,
and approximated errors (error bars).
For the latter, we only consider  
the largest source of error, the intrinsic scatter, 
and we neglect other sources of errors, as 
the uncertainty of the intrinsic scatter or
the observational error, accounted for in
our rigorous computation (which
also account for the covariance of all
sources of errors).

After binning, the decrease in the stellar mass fraction 
becomes clearer, due to the smaller error
bars of our cluster stacks.

The horizontal lines in Figure 7 and 8 
mark the Universe baryon fraction from
the five-year WMAP results in Dunkley et al. (2009).
68 \% credible intervals are shaded. This has to be
taken as an upper limit to the stellar mass
fraction, because there are other
baryons in clusters. Stars account for only about
one third at most of all baryons.

Table 2 lists derived stellar mass fractions and their errors.

\begin{table}
\caption{Stellar and gas fractions.}
\begin{tabular}{l l l l}
\hline
$\log M$ & $\log f_{stars}$ & $\log f_{gas}$ \\ 
	& $(<r_{200})$ & $(<r_{500})$ \\ 
\hline
  13.1 &                        &  $-1.18^{0.03}_{-0.03}$\\  
  13.3 &                        &  $-1.15^{0.02}_{-0.03}$\\  
  13.5 &                        &  $-1.12^{0.02}_{-0.02}$\\  
  13.7 & $-1.38^{0.06}_{-0.06}$ &  $-1.09^{0.02}_{-0.02}$\\  
  13.9 & $-1.49^{0.05}_{-0.05}$ &  $-1.06^{0.01}_{-0.02}$\\  
  14.1 & $-1.60^{0.04}_{-0.04}$ &  $-1.03^{0.01}_{-0.01}$\\  
  14.3 & $-1.71^{0.03}_{-0.03}$ &  $-1.00^{0.01}_{-0.01}$\\  
  14.5 & $-1.82^{0.03}_{-0.03}$ &  $-0.97^{0.02}_{-0.02}$\\  
  14.7 & $-1.93^{0.03}_{-0.03}$ &  $-0.94^{0.02}_{-0.02}$\\  
  14.9 & $-2.04^{0.04}_{-0.04}$ &  $-0.91^{0.03}_{-0.02}$\\  
  15.1 & 		        &  $-0.87^{0.03}_{-0.03}$\\  
\hline   	
\end{tabular}
\hfill \break  
The mass indicated in column 1 is measured within the aperture specified in the fraction 
definition.
\end{table}

\subsection{Gas mass fractions}

Vikhlinin et al. (2006)
and Sun et al. (2009) data on the fraction of matter
in the hot intergalactic gas are plotted in the right panel of
Figure 8, after converting them for minor differences in the adopted
cosmological parameters.

Masses and gas mass fractions are measured within $r_{500}$.
Because of asymmetrical errors on $f_{gas}$ 
we assume Gaussian errors on $\log f_{gas}$ 
so that our previous fitting model can be applied without any
change (apart for reading ``gas fraction" where
``stellar mass" is written). We found:

\begin{equation}
\log f_{gas} =  (0.15\pm0.03)\  (\log M500-14.5) -0.97\pm0.02
\end{equation}

with an intrinsic scatter of $0.06\pm0.01$ dex. 
The right panel of
Figure 8 shows the derived mean $f_{gas}$ vs halo mass
fit (slanted solid line) and its 68 \% uncertainty (shaded yellow
region) and the mean intrinsic scatter (dashed lines) around
the mean relation.  

Intercept and slope posteriors are Gaussian shaped, whereas
the intrinsic scatter posterior is a bit skewed, as a
Gamma function (figure not shown). Our mean relation is 
similar to Sun et al.' (2009) best fit. Their work, however, uses a similar, but
not identical, cluster sample and a different fitting model.
Our intrinsic scatter value cannot be compared with their, because
these authors, although note a scatter, do not report its value, if 
measured at all.

Similarly to stellar masses and stellar mass fractions,
gas fractions display an intrinsic
variance, i.e. intrinsic differences from cluster to cluster.
The scatter is not bounded to low mass systems, but
is apparent to all masses (see Figure 8, and Vikhlinin et al.'
2006 fig. 21), differently from
some past claims of a spread at group (low) mass only.
We emphasise that Vikhlinin et al. (2006)
and Sun et al. (2009) selected and studied a sample of
clusters and groups that, from X-ray images, appeared relaxed.
Therefore, the found spread of gas fractions at a given cluster
mass is not due to the presence
in the sample of clusters manifestly out of equilibrium (e.g.
merging).

Table 2 lists derived gas mass fractions and their errors.

Given the importance of our claim that 
intrinsic scatter is not restricted to low mass systems
only, let's look for an independent, although not equally
precise, measurement.

The existence of an intrinsic scatter at high masses is confirmed
by our analysis of gas fractions in Ettori et al. (2009).
We took their $f_{gas}$ values and errors, quoted 
at 68.3 \% level of confidence, from their Table 1. 
Their quoted errors are symmetric, and 
Gaussian distributed, because Eq. 7 in
Ettori et al. (2009) only
holds in this case. A Gaussian likelihood for $f_{gas}$ 
is mathematically impossible, because 
a Gaussian is strictly positive everywhere on all
the real axis, including impossible values for a fraction, those
outside the $[0,1]$ range. It is therefore unsurprising that
several of their quoted 68.3 \% confidence intervals include unphysical (negative) 
values for a fraction. In order to select, among their measurements,
those for which the Gaussian approximation is an acceptable
approximation,  we select from their sample only
the best measurements, defined those having a $f_{gas}$
determination with a S/N ($=f_{gas}/err$) larger than 3.
This operation removes 31 out 52 of their clusters. 
We then fit the gas fraction 
vs the cluster temperature (the mass proxy
listed in their paper) of the remaining 21 clusters, 
accounting for errors on both variables and a possible
intrinsic scatter. As in previous fitting, we re-use
the fitting model given in Appendix A, we only need to
read $f_{gas}$ and $kT$ where log stellar mass and log
halo mass is written. We found:

\begin{equation}
f_{gas} =  (0.000\pm0.004)\ (kT-8) +0.14\pm0.01
\end{equation}

with an intrinsic scatter of $0.03\pm0.01$ on $f_{gas}$.
Fig 9 shows the derived mean $f_{gas}$ vs cluster temperature
fit (solid line) and its 68 \% uncertainty (shaded yellow
region) and the mean intrinsic scatter (dashed lines) around
the mean relation. All clusters have $kT>5$ keV, i.e.
are very massive. Yet the intrinsic scatter is not zero, confirming
that the intrinsic scatter in the gas mass fraction is not reserved
to low mass clusters only.
The large majority (17 out 21) of the fitted clusters have 
an observational error smaller than the intrinsic scatter. Therefore, 
by far the largest source of uncertainty in the determination
of the average gas mass fraction of clusters (and of the
baryon fraction of the universe, derived from these measurements)
is the intrinsic scatter, and not the measurement error.

\begin{figure}
\centerline{\psfig{figure=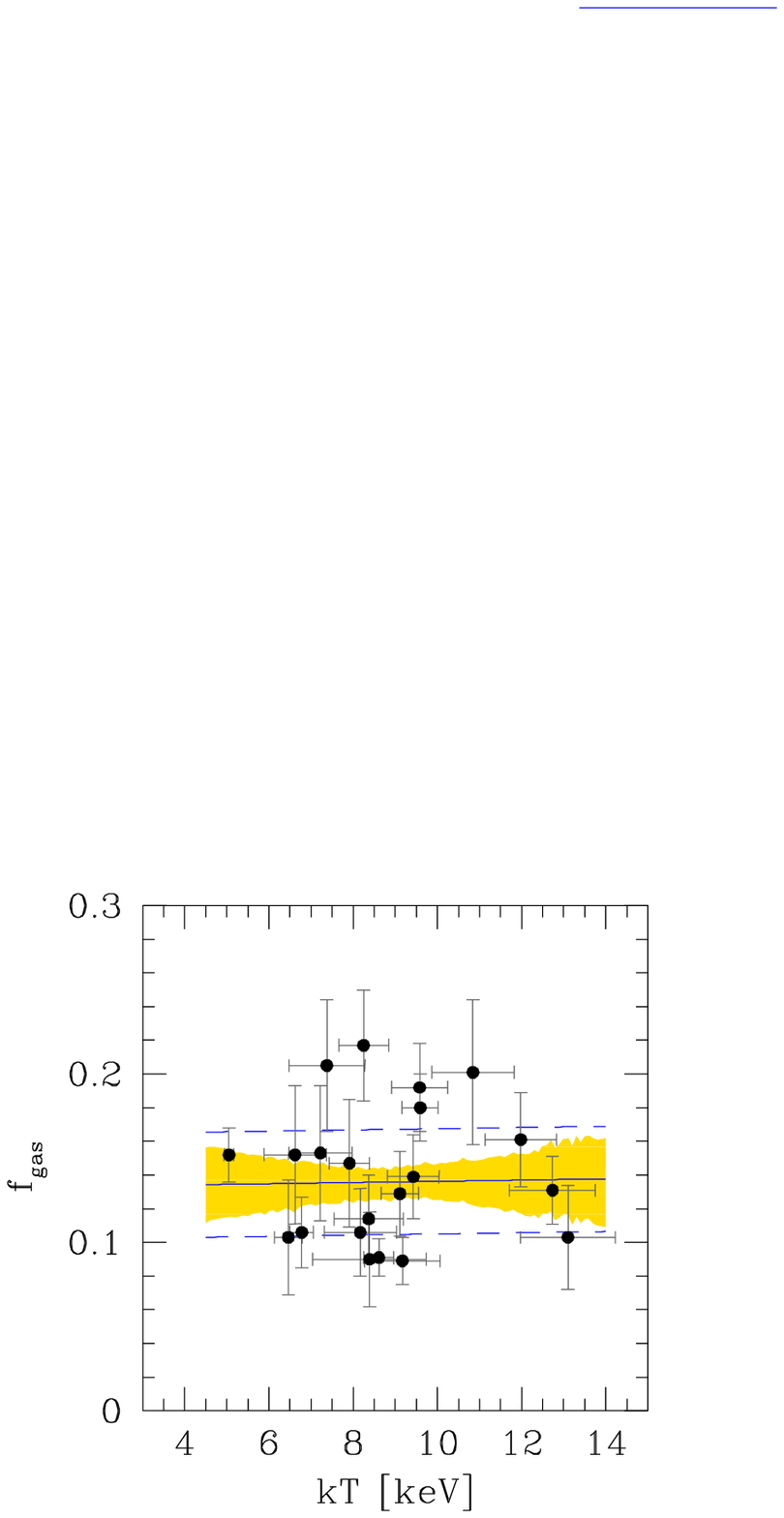,width=6truecm,clip=}}
\caption[h]{The gas mass fractions. Symbols are as in
Fig 5. Error bars on the data points (from Ettori et al. 2009) represent observed
errors for both variables. The distances between the data and the regression line is due in part to the
observational error on the data and in part to the intrinsic scatter.
}
\end{figure}

The value of the intrinsic scatter determined for the Ettori et al. (2009) 
sub-sample cannot
be easily compared with the one derived for Vikhlinin et al. (2006)
and Sun et al. (2009) for three reasons at least: first, the intrinsic
scatter is the part of the scatter not due to observational
errors. Therefore, its amplitude relies on the assumption that
observational errors are precisely measured. As mentioned,
Ettori et al. (2009) quoted errors are noisy estimate of the true
errors (see Andreon \& Hurn 2010 about how to deal with
noisy errors). Second, the scatter derived
for the Ettori et al. (2009) sample is measured on a linear
gas fraction scale, whereas the one for the Vikhlinin et al. (2006)
and Sun et al. (2009) is on a logarithmic scale (compare
the left hand side of Eq.s 4 and 5), and a Gaussian
scatter on one scale does not translate on a Gaussian scatter
on the other scale.  Finally, the scatter measured with the
Ettori et al. (2009) data is at a given temperature, not at
a given mass, as the one measured with Vikhlinin et al. (2006)
and Sun et al. (2009) data. If the above caveats are ignored, 
then one find a good
qualitative agreement between the two determinations of the
intrinsic scatter after an approximate conversion to a common scale.
 
\begin{figure*}
\centerline{\psfig{figure=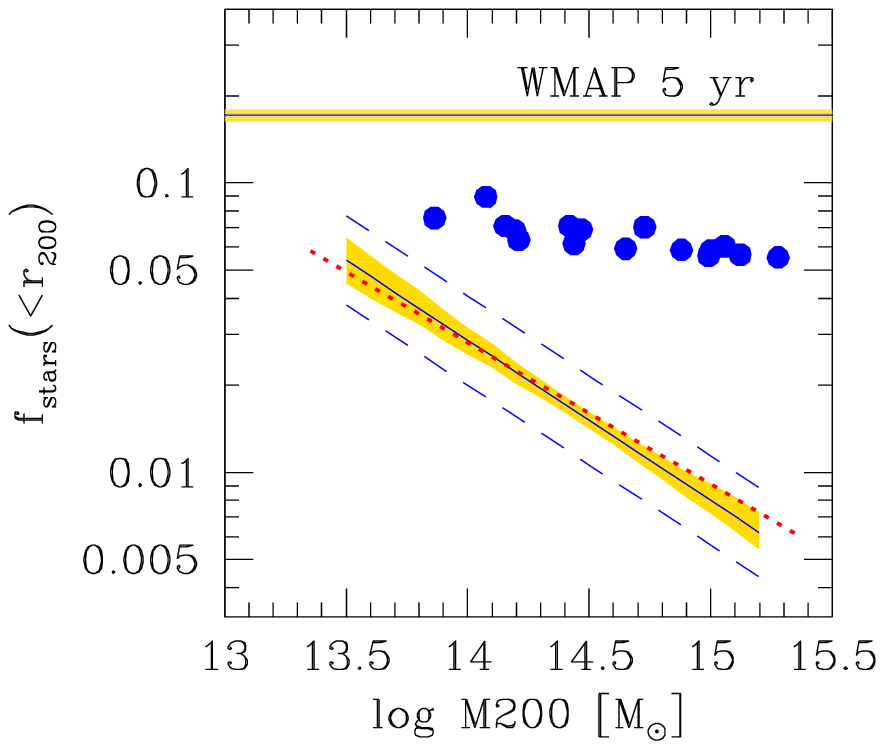,width=8truecm,clip=}
\psfig{figure=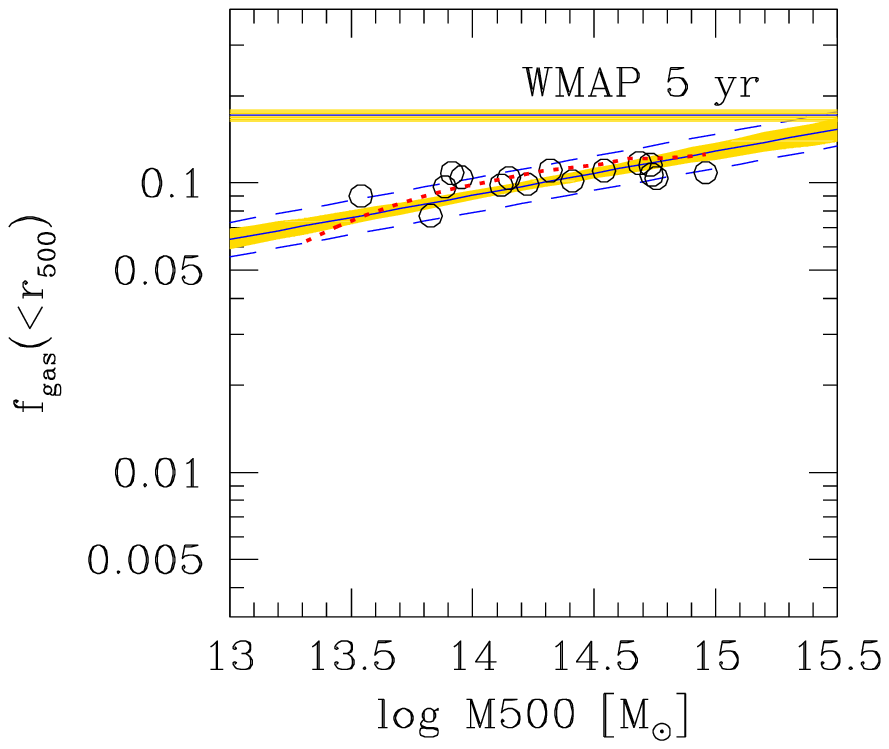,width=8truecm,clip=}}
\caption[h]{Comparison between observed and theoretical stellar (left panel) and gas
(right panel) fractions. 
Solid lines and shaded regions mark our
observational constraints on the stellar and gas mass
fractions. Open and closed
points mark theoretical gas and stellar mass fractions, respectively, 
observed in gasdynamical simulation 
with cooling and star formation of Kravtsov et al. (2005)
and Nagai et al. (2007).
The dotted red lines are the predictions of 
the model by Bode et al. (2009).}
\end{figure*}
 
The existence of an intrinsic scatter in $f_{gas}$ 
(both at a given mass and at a given Temperature)
must surprise all those who believe
that, having clusters collected material from
a large region, their content should be representative
of the mean matter content of the universe (White et al. 
1993). The measured intrinsic scatter implies that when taken 
{\it individually} the regions in which 
clusters and groups collected matter, a few tens of Mpc, are yet
not representative, in terms of gas content and therefore
in the baryon content (at large masses $f_{gas}$ is the largest
baryon contributor) of the mean matter content
of the Universe, i.e. each region has a gas, and thus baryon, content
that differs from the average by more than the observational error.
The existence of an
intrinsic scatter on $f_{gas}$ does not preclude the use of the gas 
or baryon mass
fraction as determined in clusters
for cosmological tests, it only decrease its
efficiency (a larger sample is required to achieve the
same precision), and oblige us to address
selection effects, i.e. to inquire if studied clusters are 
representative, in terms of $f_{gas}$, to the population present in
the Universe, or are a biased subsample. Therefore, cosmological
constraints derived from $f_{gas}$ ignoring  intrinsic scatter and
$f_{gas}$ selection function (e.g. Ettori et al. 2009) are optimistically 
estimated, and, perhaps, biased.

\subsection{Stellar and gas mass fractions}

The comparison of right and left panel of Figure 8 
shows that the halo mass at which stars and gas contribute equally
to the total halo baryonic content is near $10^{13.5}$ solar
masses (roughly, $M200 \approx 1.5 \ M500$).

Figure 10 compare our observational constraints
and theoretical predictions on
the stellar (left panel) and gas (right panel) mass fraction.
We consider gasdynamic simulations of Kravtsov et al. (2005)
and Nagai et al. (2007).
These simulations are performed in a (simulated) universe
with a slightly too low (compared to WMAP5) baryon fraction
($0.14$ vs $0.17$). Therefore we revise upward fractions
derived in simulations  
by $0.17/0.14$. Stellar mass fractions (Kravtsov 2009, priv. comm.)
are measured at $r_{200}$
in the simulations, as for data. Gas mass fraction for the 
very same simulations are taken from Nagai, Vikhlinin \& Kravtsov
(2007), and are measured within $r_{500}$, as for data.
Gas fractions have been revised upward by  $0.17/0.14$, for
the same reason as stellar fractions.
The predicted fraction of matter in gas (open points, right panel) is
near to the observed gas mass fraction, as commented in
Kravtsov et al. (2009) and Nagai et al. (2007). Note in 
particular that simulated
clusters also displays a spread of gas and stellar mass fractions at a given
halo mass, as real clusters.

As remarked by Kravtsov et al. (2009) and Gonzalez et al. (2007), 
the predicted stellar mass fraction (close points, left panel)
is off from the observed one (slanted solid line), 
and more evidently so in our paper than in previous works,
given our more precise observational determination.
Both the slope and the intercept of the simulations are 
in flagrant disagreement with the observed values. Gasdynamic
simulations with star formation of Fabjan et al. (2009) show
a very similar behaviour to Kravtsov et al. (2009) and Nagai et al. (2007)
simulations:  the temperature-mass scaling and the gas mass fraction can 
only be reproduced if star formation is allowed, but in such
case the predicted stellar mass fraction (similar in the two
sets of simulations) is off.

The mismatch between predicted and observed stellar mass fractions
is not of secondary importance in the cluster
model because of the strict interplay of the stellar and
gas components. First, if less stars need to
be formed, then more gas is left, and thus the current
agreement between predicted and observed gas mass fractions is 
corrupted.  Second,
the gas component responds to the feedback of the stellar component 
(e.g. Nagai et al 2007; Fabjan et al. 2009). 
If the stellar part of the model
have to be altered, changes on the gas predictions 
(e.g. X-ray scaling relations) occur. Unfortunately,  
the agreement of gas-related quantities, as the temperature-mass
scaling, holds in current models  for wrong 
stellar content predictions.
Therefore, observations on the stellar mass fraction
presented in this paper give a challenging
constraint to theories of cluster formation. 

Bode et al. (2009) presented a semi-analytical cluster model, i.e.
they inserted a number of recipes on an N-body simulation\footnote{A number
of recipes are also inserted in to gasdynamic simulations.}. They
concluded that if the stellar mass fraction has a logarithmic
slope of $-0.49$, then there is no need of a supplementary
feedback,  i.e.  in addition 
to the stellar one,
to match the gas mass fraction and X-ray scale relations
(temperature-mass, $Y_x$-mass).
Our observed stellar mass fraction has a logarithmic slope of
$-0.55\pm0.08$ is consistent with the slope required 
to avoid supplementary feedback in the Bode et al (2009) model.

In the Bode et al. (2009)
model, this is a real
model prediction, it has not adjusted to match previous 
observational data on the stellar mass fraction slope. 
At the contrary, the remaining model parameters commented
below have been adjusted to fit observations, 
reducing the significance of the agreement between
``predicted" and observed values.
The intercept of the model stellar mass
fraction vs mass, kept fixed by Bode et al. (2009) to the
Lin et al. (2003) value, also agrees with our observational
determination: the Bode et al. (2009) model 
stellar mass fraction (dotted red
line, left panel) is fully enclosed in the 68 \% confidence band of
our observational determination, i.e. it is in remarkable
agreement. Similarly, the Bode et al. (2009) model
gas mass fraction (dotted red line, right panel) is in reasonable
agreement with our summary of
observational data.
If the simple Bode et al. (2009) cluster model
is not an over-simplified description of true existing clusters, 
then the slope of the stellar mass fraction vs halo mass we
determine in this paper implies that AGN feedback is not needed, at least
to reproduce X-ray scaling relations and stellar and gas mass fractions.

A couple of technical points are worth mention:
in the Bode et al. (2009) model, the stellar mass fraction 
is determined within the virial radius, as our observational
determination is, but is parametrised as a
function of $M(<r_{500})$, that we converted to 
$M200=M(<r_{200})$,
assuming a NFW profile and a concentration of 3. Instead, gas mass 
fractions are computed within $r_{500}$ both for the model and
for the data.

\subsection{Stellar + gas mass fraction}

In order to add the stellar and gas mass fractions to get the
total fraction of baryons we need to 
address some issues.

First, stellar and gas mass fractions are measured inside different 
reference radii ($r_{200}$ and $r_{500}$). Simulations
shows that the region inside $r_{500}$ is depleted, if any, by a small
and poorly determined amount of the order
of 2 to 10 \% (Ettori et al. 2006, Kravtov et al. 2005). We adopt
a 5 \% correction with a $3 \sigma$ error of 6 \%.

Second, we need to convert the fit in Eq. 3 from a fit vs $M500$
to a fit vs $M200$. The mass conversion is performed assuming
a NFW of concentration 3. Any scatter of the concentration at a
given mass or any change of the mean concentration with mass 
has little effect on this conversion, because Eq. 3 is 
the mean relation, and it is linear and shallow.

Third, the gas and stellar masses
measurements are assumed to be independent (which is true), and 
the intrinsic scatter of gas and stellar mass fractions
around the mean are assumed to be 
unrelated each other (which
is unknown given the available data).

\begin{figure}
\psfig{figure=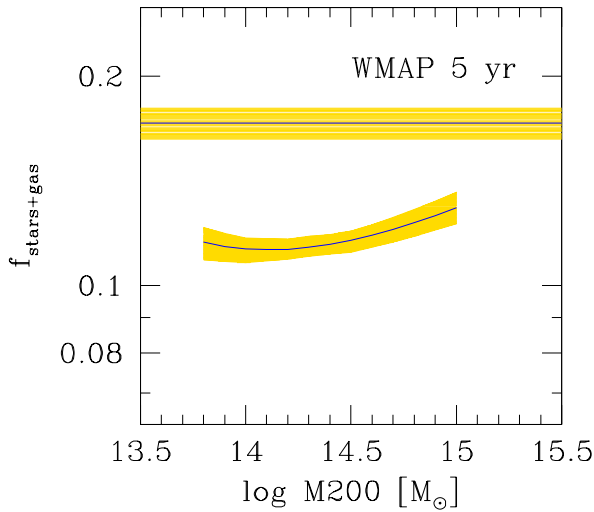,width=8truecm,clip=}
\caption[h]{Baryon fraction.
The solid curve and shaded region mark the mean 
the observationally measured $f_{stars}+f_{gas}$
and its (rigorous) 68 \% confidence
error. The WMAP 5 yr baryon fraction value and 
error (Dunkley et al. 2009) are also marked.}
\end{figure}

Figure 11 displays the total fraction in baryons, $f_{stars+gas}$,
as a function of cluster mass, in the range where both $f_{stars}$ and 
$f_{gas}$ are both constrained
by the data, $13.7 \le \log M200 \le 15.0$ solar masses. 
These values and errors come from the fit
on individual data points in the stellar vs total mass 
plane and in gas mass fraction vs total mass, with the mentioned (minor)
corrections. The variety of cluster properties at a
given mass is fully accounted for by our derivation of $f_{stars+gas}$.

Two points are striking in Figure 11:
a) we observe an almost constant baryon fraction in
the studied mass range, the increase of the gas mass fraction
being approximatively compensated by the decrease of the stellar
mass fraction; and
b) WMAP-derived baryon fraction differs from our
estimate by about $6$ sigma's.

Readers interested in inferring the
values of cosmological parameters from our measured
baryon fraction should remember that our own
determination of the baryon fraction has 
been derived for an assumed set of cosmological
parameters (listed at the end of the introduction), 
instead of being marginalised over the uncertainty of
cosmological parameters. The latter operation matters for
estimates of cosmological parameters, which is, however,
beyond the scope of the present paper.

How can one reconcile the observed value of the
baryon fraction in clusters with the larger value derived from WMAP?

We explore few possibilities in turn.

First, might a large bias on $f_{stars}$ be present?
It is hard to be accommodated, because 
in order to boost the stellar mass fraction one need:

a) that $L_{tot}$ is largely underestimated. This is only possible
below the observed range of galaxy luminosities (we reach 
$-19.3<M_r<-15.7$ mag, depending on redshift) and requires 
that the luminosity function keeps a 
diverging slope (i.e. $\alpha \la -2$) for a large magnitude
range below the one considered in this work. 
However, clusters with very
deep observations do not display such a feature, down to
$M_V\sim-11$ (e.g. Sandage, Binggeli, \& Tammann 1985;
Andreon \& Cuillandre 2002; Andreon et al. 2006; Bou\'e et al. 2008,
etc); or

b) that the observed (and adopted) $M/L$ conversion is biased; 

c) that the intracluster light (that we emphasise not to
include the light from galaxy outer halos,
from the BCG and from undetected galaxies, see Sec 3.2) is about 5 mag arcsec$^{-2}$ brighter
that the value measured in Zibetti et al. (2005). This number has been
derived basically reverting the performed operations to derive
stellar masses: we first compute the missing baryon
fraction (WMAP $f_{baryon}$ value minus $f_{star+gas}$), we multiply it by the
cluster mass within $r_{200}$ to derive missing stellar masses.
We then convert the latter in luminosities with
the assumed $M/L$ value, and project
them in the plane of sky (i.e. converting back from
values within spheres to within cylinders). Finally,
mean brightness is derived from the luminosity value, accounting for the
cluster size (i.e. adding $2.5 \log \pi r^2_{200}$, with
radii in arcsec). 
The missing mass implies a mean (within $r_{200}$
and averaged over the sample of 52 clusters) surface brightness
of $\mu_r \sim 26.5$ mag arcsec$^{-2}$. The latter value is, as mentioned,
fairly too bright to match the observational value.

Even if a large bias on $f_{stars}$ is there (which is implausible)
for a still un-identified reason, this leaves untouched 
the disagreement with the WMAP-derived value at 
high cluster masses, where the stellar contribution is minor.

Second, $f_{gas}$ estimates might be systematically low. Systematic
biases,  in addition to depletion already accounted for, 
of the gas mass fraction are discussed in Ettori et al. (2009),
Allen et al. (2008) and reference therein. Our reading of these
papers is that the gas mass fraction is free from
important unknown systematics, otherwise these
authors would not attempt to constraint cosmological parameters
using it. Therefore, we are tempted to
exclude systematics on $f_{gas}$, although, of course,
direct measurements of $f_{gas}$ within $r_{200}$ would be
preferable. We only known a single $f_{gas}$ measurement at $r_{200}$,
by George et al. (2009). They found
a fraction $0.02\pm0.02$ higher at $r_{200}$ than at $r_{500}$,
in agreement with our assumed correction.

Third, we might miss some other sources of baryons. Fukugita, Hogan
\& Peebles (1998) explored this issue, and
concluded that stars and the hot intergalactic gas contain
the large majority of baryons.

Therefore, we are obliged to consider the possibility that
the WMAP baryon fraction is wrong in some sense.
The value derived by WMAP is the value of the baryon
content assuming that it is universal,
i.e. equal to an unique value everywhere without any scatter
or dependency with anything (say halo mass). This paper
clearly shown that
$f_{stars}$ and $f_{gas}$ are both not-universal:
these fractions have a mass dependency. Furthermore,
there is also a spread of both the stellar and gas mass
fraction at a given mass. 
In absence of a fine tuning
between the two fractions (a stellar
mass excess being compensated by a gas deficit, which is hard
to obtain given their widely different contributions at the
two ends of the halo mass function), 
$f_{baryons}=f_{stars}+f_{gas}$ should display an intrinsic scatter.
The hypothesis of a possible not universal baryon fraction, although
surprising, is not totally new and has been
already proposed by Holder, Nollett, \& van Engelen (2009).
It would be interesting to known what
would be the CMB-derived constraint on the baryon fraction if
it was allowed to display a  
variance. 

It is also possible that the baryon fraction is larger than the
WMAP value at locations that we have not sampled, 
halos with masses
lower than $10^{13.5}$ solar masses and outskirts of clusters.
For the latter, there is some evidence (e.g. Rines et al. 2004). 
A larger-than-WMAP
baryon fraction at these locations might compensate the 
lower-than-WMAP
value in the studied (portion of) clusters and groups. 
This possibility assumes a non-universality of the baryon fraction.
In such case, WMAP-derived baryon fraction might require a 
new derivation, under this less restrictive hypothesis.

Finally, McCarthy et al. (2007)  suggested
that the problem may sit in an underestimate of the 
denominator of the WMAP baryon fraction, i.e. in $\Omega_m$.

Although we have no solved the baryon discrepancy,
we can exclude almost for sure that the stellar fraction
is responsible for the difference between cluster and global
baryon fraction and we identify possible points that
require investigation.

\subsection{Comparison with previous works}

Our results are in qualitatively agreement with previous results 
(Gonzalez et al. 2007; Giodini et al. 2009), displaying an
offset between the measured total baryon fraction and the WMAP value, and
a decreasing stellar mass fraction with increasing mass. Some
specific results might instead differ from some published
results, for example, we do not confirm the Giodini et al. (2009) claim 
that the baryonic fraction increases with mass. However,
our statements are fundamentally different from those published in other
works.

We have already discussed, in sec 3.5, about the advantages
of an accurate analysis of the data, of 
studying a sample of truly existing clusters 
located in a narrow range of
redshift and having
individually measured, and large, reference radii. 
We continue along the same line by noting
that the cluster mass enters in the stellar mass fraction
(it is at the denominator of the fraction), and 
thus is certainly an advantage to study clusters with known and
precisely measured masses, instead of those with noisy or
unknown masses.
Our sample has accurate cluster masses derived
under parsimonious hypothesis, that does not require
the cluster being in dynamical or hydrostatic equilibrium,
from the caustic analysis
of Rines \& Diaferio (2006) of about 208 galaxy members
per cluster, on average. 
Instead, Gonzalez et al. (2007) masses are derived from
velocity dispersion computed on small samples and 
with Beers et al (1990) estimators. These velocity
dispersions (and masses) have low reliability (Andreon et al. 2008;
Andreon 2009;  Gal et al. 2008). Giodini
et al. (2009) assume that mass is proportional to a
poorly estimated X-ray
luminosity without any scatter, when instead the two quantities
display a large scatter (e.g. Stanek et al. 2006; Vikhinin et al.
2009; Andreon \& Hurn 2010), it is just enough to remember
the existence of cool-core clusters.

A second key difference between our and other works lays
in performing an analysis that do not contradict the
expected and observed spread in cluster properties
at a given mass. Galaxy groups and clusters
are the result of the assembly history of dark matter
halos, and also shaped by star formation processes affecting
the gas. These physical processes (and possibly other) 
lead to multivariate outcomes 
and produce an intrinsic spread in the distribution of the
observed properties of groups and clusters, a spread
that is readily apparent in any stellar or gas mass fraction vs 
cluster mass plot, such as our Figure 7 or 8. The spread manifest
itself also as a variance of concentrations (and thus $r_{200}$) 
at a given cluster mass. Therefore, it
is of paramount importance to account for the variance of
cluster properties at a given cluster mass. Previous
analysis fail to account for the above, having 
all assumed instead that clusters are identically equal at a 
fixed mass (or mass proxy). For example, 
Giodini et al. (2009)
assume that all clusters of a given
X-ray luminosity  have the same size. 
Our analysis allows a variance in the cluster properties
at a given halo mass. Finally, we have already noted
that our analysis avoid the use of fitting 
methods in conditions where they must not be
used.

In spite of our accounting of a larger number of error terms, we
are able to reject the WMAP value at ``much more sigma's" than
previous works, $6$ vs $3.2$ (Gonzalez et al. 2007) or $\sim 5$ 
(Giodini et al. 2009). This is the result of 
a different choice of the data and 
cluster sample: we choose clusters with accurate masses,
good photometry and low galaxy background contribution, i.e.
nearby clusters with caustic masses.

\section{Summary}

We analysed a sample of 52 clusters with precise
and hypothesis-parsimonious measurements of mass, derived from
caustics based on about 208 member velocities per cluster
on average, and with measured $r_{200}$ values.
We found that low mass clusters and groups
are not simple scaled-down version of their massive cousins
in terms of stellar content: lighter clusters have more
stars per unit cluster mass. The same analysis also
shows that the stellar content of clusters displays an
intrinsic spread at a given cluster mass, i.e. clusters are not similar
each other in the amount of stars they contain, not even at a fixed 
cluster mass. 
The amplitude of the spread in stellar mass, at a fixed
cluster mass, is $0.15\pm0.02$ dex. 
The stellar mass fraction 
depends on halo mass with (logarithmic) slope  $-0.55\pm0.08$.
These results are confirmed by adopting
masses derived from velocity dispersion.

The intrinsic scatter at a fixed cluster mass we determine for gas 
mass fractions taken from literature, is 
smaller, $0.06\pm0.01$ dex. The intrinsic spread is not
restricted to low mass systems only, but extend
to massive systems. Since the studied systems look relaxed
in X-ray images, the found spread is not due the presence
in the sample of clusters manifestly out of equilibrium
(e.g. merging). The non-zero intrinsic
scatter of the gas mass fraction decreases the efficiency 
of $f_{gas}$ for cosmological studies, and asks to inquire about
whether studied samples are representative, in terms of 
$f_{gas}$, to the population of clusters in the Universe.

The intrinsic scatter in both the stellar 
and gas mass fraction is a distinctive signature that 
when taken individually the regions in which 
clusters and groups collected matter are yet
not representative, in terms of stellar and gas content and therefore
in the baryon content, of the mean matter content
of the Universe.

The observed stellar mass fraction values are in marked
disagreement with gasdynamics simulation with cooling and
star formation of clusters and groups. Instead,
amplitude and cluster mass dependency
of observed stellar mass fraction are those requested
not to need any AGN feedback to describe X-ray scale relations
and gas and stellar mass fractions in simple semi-analytic cluster models.

By adding the stellar and gas masses, or, more precisely speaking,
by fitting both them and accounting for
the intrinsic variance of both quantities,
we found that the baryon fraction is fairly
constant for clusters and groups with $13.7<\log M200 < 15.0$ solar masses and
it is offset from the WMAP-derived value by about $6$ sigmas. The 
offset is unlikely to be due to an underestimate of the
stellar mass fraction and could be related to the possible
non-universality of the baryon fraction, pointed out by
our measurements of the intrinsic scatter. 

Our analysis is the first that does not assume
that clusters are identically equal at a given halo
mass and it is also more accurate in many aspects than
previous works. The data and code used for
the stochastic computation are distributed with the paper. 

\section*{Acknowledgements}

We thanks Andrey Kravtsov and Paul Bode for giving us their 
theoretical predictions in electronic format. The author
greatly benefitted by comments on this draft from Stefano Ettori,
Fabio Gastaldello, Stefania Giodini, Anthony Gonzalez, Paolo Tozzi
and the referee.
For the standard SDSS and NED acknowledgements see: 
http://www.sdss.org/dr6/coverage/credits.html and
http://nedwww.ipac.caltech.edu/.

\appendix

\section{Model listing and coding}

In this section we give the listing of the full model, and
its coding in JAGS.

Observed values of halo and stellar mass 
($obslgM200$ and $obslgM_{\star}$, respectively) have a
Gaussian likelihood. 

\begin{eqnarray}
obslgM200_i &\sim& \mathcal{N}(lgM200_i,errlgM200^2_i) \\
obslgM_{\star,i} &\sim& \mathcal{N}(lgM_{\star,i},errlgM_{\star,i}^2) 
\end{eqnarray}

The tilde symbol reads ``is drawn from", and the symbol $\mathcal{N}(y,\sigma^2)$
denotes a Gaussian centered on $y$ with variance $\sigma^2$.

True values of stellar and halo mass are
linearly related (on a log scale), with an intrinsic scatter
$\sigma_{scat}$.

\begin{eqnarray}
z_i &=& \alpha+12.5+\beta \ (lgM200_i-14.5) \\ 
lgM_{\star,_i} &\sim& \mathcal{N}(z_i, \sigma^2_{scatt}) 
\end{eqnarray}

Masses are re-centred, purely for computational advantages in
the MCMC algorithm used to fit the model (it speeds up
convergence, improves chain mixing, etc.).
Please note that the relation is between true values, not
between observed values, which may be biased. 

Uniform priors are taken: the halo mass, $obslgM200$, has
a strictly uniform prior; the intercept, $\alpha$ 
has a zero mean Gaussian 
with very large variance; the slope, $\beta$, 
has a uniform prior on
the angle (that becomes a Student $t$ distribution for the
angular coefficient), because we do not want that cluster
properties depends on astronomer rules to measure
angles (see Andreon 2006, 
and Andreon \& Hurn 2010 for a discussion); 
$1/\sigma^2_{scat}$ has a Gamma distribution
with small values for
its parameters, as in 
Andreon \& Hurn (2010). This has the welcome property that the
intrinsic scatter variable is positively defined, 
as the intrinsic scatter is. In symbols:

\begin{eqnarray}
lgM200_i &\sim& \mathcal{U}(0,500) \\ 
\alpha &\sim& \mathcal{N}(0.0,10^4) \\ 
\beta &\sim& t_1 \\
1/\sigma^2_{scatt} &\sim& \Gamma(10^{-5},10^{-5}) 
\end{eqnarray}

Our model makes weaker assumptions 
than other models adopted in previous analysis, 
and plainly states what is actually also assumed
by other models (e.g. the conditional independence and
the Gaussian nature of the likelihood), also 
removing approximations adopted in other approaches.

For the stochastic computation and for building the statistical
model we use Just Another Gibb Sampler 
(JAGS\footnote{http://calvin.iarc.fr/$\sim$martyn/software/jags/}, 
Plummer 2008).
JAGS, following BUGS (Spiegelhalter et al. 1995), uses 
precisions, $\tau = 1/\sigma^2$, in place of variances $\sigma^2$. 
The arrow symbol reads ``take the value of". 
Normal, t, and Gauss distributions are indicated with
{\texttt{dpois, dt, dgamma}}, respectively.

\begin{verbatim}
model 
{
for (i in 1:length(obslgMstar)) {
  obslgM200[i] ~ dnorm(lgM200[i],tau.lgM200[i])
  lgM200[i] ~ dunif(0,500)
  
  obslgMstar[i] ~ dnorm(lgMstar[i],tau.lgMstar[i])
  z[i] <- alpha+12.5+beta*(lgM200[i]-14.5)
  lgMstar[i] ~ dnorm(z[i], prec.intrscat)
  } 
intrscat <- 1/sqrt(prec.intrscat)
prec.intrscat ~ dgamma(1.0E-5,1.0E-5)
alpha ~ dnorm(0.0,1.0E-4)
beta ~ dt(0,1,1)
}
\end{verbatim}

The code above, which is an almost literal
translation of equations A1 to A8, is only about 10 lines 
long in total, about two orders of magnitude
shorter than any previous implementation of a regression model 
(e.g. Kelly 2007, Andreon 2006, Akritas \& Bershady 1996). 
The model is quite general,
and applies to every quantities linearly related, with Gaussian
errors, and with an intrinsic scatter. 
For example, in this paper the same model (and program) 
has been used for the $f_{gas}$ vs mass, and $f_{gas}$ vs 
temperature scaling. The same fitting model can be used for,
say, the $L_x-\sigma_v$, 
richness--mass and halo occupation number vs mass scalings.

\end{document}